# Rupture-Repair Cycles in Regenerating *Hydra* Tissues

Oded Agam[1] and Erez Braun[2]

[1]The Racah Institute of Physics, Edmond J. Safra Campus, The Hebrew University of Jerusalem, Jerusalem 9190401, Israel.

[2] Department of Physics and Network Biology Research Laboratories, Technion-Israel Institute of Technology, Haifa 32000, Israel.

**Abstract**

Destructive mechanical breakdowns and fractures are ubiquitous events in driven physical matter; living tissues, by contrast, can rupture repeatedly while restoring integrity. Here we study rupture–repair interplay in regenerating *Hydra* tissues, which cycle through osmotic inflation, pressure release by rupture, and resealing. We utilize bright-field imaging of the tissue's projected area as a readout of the rupture magnitude before it is arrested. Analyzing these event statistics, we find that the tail of the area-drop distribution is controlled by $Ca^{2+}$-dependent repair efficiency. When the $Ca^{2+}$ response is weakened, either by partially blocking gap-junctions mediating the intercellular communication, or by inhibiting stretch-activated $Ca^{2+}$ channels, the actomyosin force that arrests the rupture process is delayed or reduced. Under these conditions, rare large pressure releases become more likely, and the tail of the distribution crosses over from an exponential behavior, exhibiting a characteristic scale, to a power-law one consistent with a critical-like regime reflecting intermittent rupture propagation. These results identify mechanically evoked $Ca^{2+}$ activity as a control axis linking repair to rupture statistics in a living tissue. It supports a picture of rupture front advancing by stick-slip-like dynamics as it encounters a heterogeneous mechanical landscape, akin to failure-front propagation in disordered materials.

**Introduction**

Tissue repair and wound healing are indispensable for the persistence of multicellular life, yet they must unfold through tightly coordinated processes spanning multiple organizational levels[1]. Repair frequently entails the rapid mobilization and proliferation of cells, and the generation of active forces to restore tissue cohesion[2, 3]. This raises a central question: how can tissues reliably build and rebuild a functional form under continuously changing, active conditions, including those that drive them to the brink of mechanical failure? Addressing this question can reveal general principles that govern animal morphogenesis and regeneration, and is of prime importance for medical applications.

A common signature of driven systems near failure is intermittent dynamics in which gradual loading is interrupted by abrupt, localized events[4, 5]. Tissue tearing and resealing can be viewed through this lens, with active stresses driving the tissue toward rupture and coordinated responses restoring cohesion. This perspective motivates a quantitative approach that quantifies the distributions of rupture and repair events and examines how these statistics vary under controlled perturbations. By doing so, we can connect cellular-scale activity with tissue-level mechanics and reveal key features of the repair mechanism.

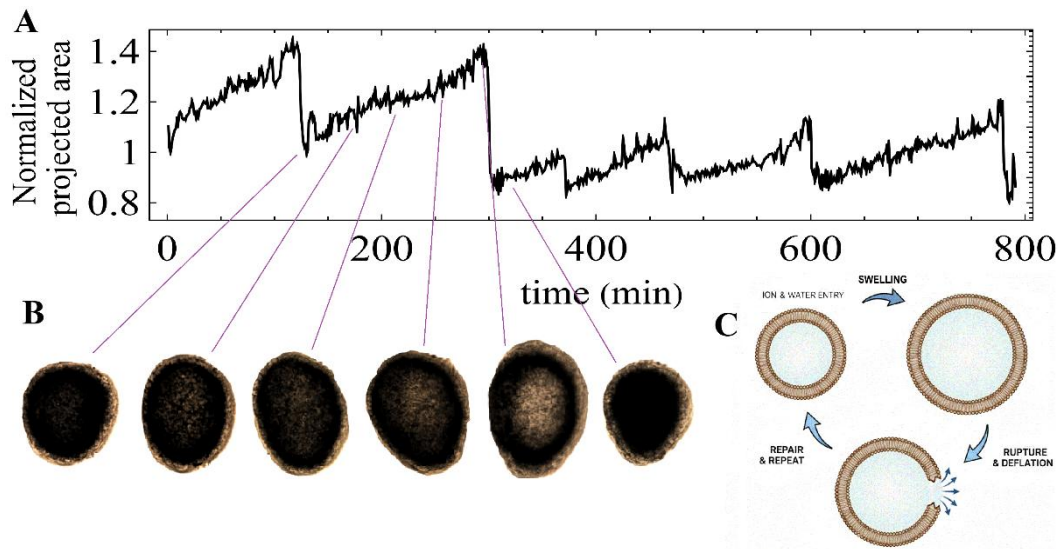

**Figure 1 | inflation–rupture–repair cycle and area-drop events.** (A) A representative sample out of a long-time trace, of the normalized projected area of a regenerating *Hydra* tissue fragment, showing gradual increases punctuated by abrupt drops. The projected area is normalized by its mean over the full-time trace. Large drops correspond to pressure-release (rupture) events, and the gradual segments reflect re-inflation. (B) Example of bright-field snapshots from the same recording at the time points indicated in (A), illustrating gradual swelling and rapid deflation during an area-drop event. (C) Schematic of the cycle: an osmotic pressure gradient drives ions and water into the lumen, leading to swelling and internal hydraulic pressure buildup; this pressure is released by rupture and deflation; the tissue then repairs and reseals, enabling the cycle to repeat.

*Hydra* regeneration from a tissue fragment offers a uniquely tractable setting for studying rupture and the rapid restoration of tissue cohesion under relatively simple geometric and mechanical constraints[6-10]. During the early stages of regeneration, when cell proliferation is relatively rare[11, 12], tissue dynamics are governed by the interplay between rising internal pressure and active actomyosin contractility[10, 13-15]. The resulting mechanical loading repeatedly drives the tissue to rupture which then rapidly reseals, generating a natural sequence of well-defined failure events[16]. These intrinsic, well-characterized ruptures, together with the simplified tissue architecture and the dominance of a single force generation source- the actomyosin system[10, 17, 18]- make *Hydra* an ideal system for probing the physics of tissue repair in a living, developing organism.

During regeneration, a small fragment excised from the body of a mature *Hydra* folds into a hollow shell made of a bilayer epithelial tissue that encloses a fluid-filled lumen[10]. The sealed tissue shell ensures the maintenance of an ionic reservoir in the lumen, essential for cellular homeostasis in the freshwater low-ionic content environment[13, 14, 19]. Osmotically driven influx gradually inflates the lumen and builds internal pressure, mechanically loading the epithelial bilayer. Once the load exceeds the tissue's local tolerance, a transient opening forms and the lumen pressure is relieved on a much shorter timescale. The opening then reseals, rapidly restoring cohesion. We operationally define resealing as the time at which lumen pressure begins to recover; yet, full restoration of the tissue's structure may continue beyond this point[20]. As defined here, resealing dynamics can be quantified from bright-field images by using the projected tissue area as a proxy for enclosed volume, producing a characteristic sawtooth trace: slow swelling punctuated by abrupt drops that repeat over successive cycles (Fig. 1).

A simple interpretation of repeated inflation and rupture would be a well-defined failure threshold that sets a characteristic event scale. The regenerating tissue, however, is intrinsically heterogeneous, and its resistance to failure varies in space and evolves in time with cellular state, cell-cell adhesion, cytoskeletal organization, tissue geometry, and extracellular matrix. Tissue failure is therefore governed by a distributed landscape of local thresholds rather than a single global one. Accordingly, rupture dynamics can be viewed within the broader class of slowly driven failure processes with burst-like stress release, exemplified by earthquakes[21]. In disordered media, stress is often released through the advance of failure fronts propagating through a landscape of weak spots and barriers, rather than by the smooth growth of a single, well-defined crack[4, 22]. Front motion is, therefore, typically intermittent, with arrests at local barriers punctuated by rapid advances, so that a single event can comprise multiple bursts spanning a range of timescales[4, 23, 24]. By analogy, a rupture episode in the *Hydra*'s tissue needs not unfold smoothly but can consist of a sequence of arrests and rapid expansions shaped by spatially varying mechanical constraints. This naturally leads to cycle-to-cycle variations in rupture extent and timing, calling for a statistical characterization of rupture events. However, rupture does not unfold in isolation. Local changes in stretching and stress that accompany pressure release engage repair activity and restore tissue cohesion on about a minute timescale. This healing response relies on active forces and mechanosensitive feedback that act to arrest the opening and promote resealing. Consequently, perturbing the repair response provides an experimental control over resealing efficiency, exposing rupture events that are either rapidly terminated or prolonged.

A direct and testable consequence of this picture is encoded in the event statistics[4, 5, 22, 25]. Although rupture propagation is not directly resolved, its signatures can be inferred from the statistics of pressure-release events which appear in our measured time series as sharp drops in the tissue's projected area. When resealing is efficient and rupture growth is rapidly arrested, event sizes are expected to be characterized by a typical scale. When resealing efficiency is compromised, intermittent rupture growth leaves a stronger imprint on the statistics, increasing the weight of rare, large area drops and modifying the tail of the area-drop distribution. Controlled perturbations of the repair machinery, therefore, provide a practical tool for probing the nature of rupture propagation in the tissue. If resealing is accomplished by actomyosin-driven contractile forces, as suggested by the prominent involvement of actomyosin in the regeneration process[10, 17, 18, 20, 26], then $Ca^{2+}$-targeted interventions provide a natural route to perturb the process. Indeed, as shown before, $Ca^{2+}$ plays a crucial role in *Hydra* morphogenesis[6, 27-30]. Accordingly, modulating $Ca^{2+}$ activity and responsiveness offer an effective experimental handle to control and probe the rupture-resealing dynamics.

Our central finding is that $Ca^{2+}$-linked repair dynamics, indeed, tune the tail of the area-drop distribution. Under baseline conditions, rupture events are rapidly terminated by contractile closure, yielding an exponential tail with a characteristic event scale. Weakening the $Ca^{2+}$ response, either by partially blocking gap-junctions mediating the intercellular communication, or by inhibiting stretch-activated $Ca^{2+}$ channels, modifies the statistics: rare large releases become more likely, and the distribution crosses over to a power-law tail, consistent with a critical-like regime of stress release[25]. Enhancing $Ca^{2+}$ activity with an applied moderate-amplitude electric field, which does not inhibit the regeneration process, produces the opposite trend, suppressing large releases and retaining an exponential tail while reducing the characteristic scale. Taken together, these perturbations support a picture in which rupture growth is intermittent and shaped by the tissue's heterogeneous mechanical landscape. They also identify the mechanically evoked Ca²⁺ response as a control axis that regulates resealing efficiency and tunes the system toward or away from criticality.

**Results**

Small tissue fragments of a few hundred cells were excised from the body of a mature *Hydra* and allowed to fold and seal into closed, hollow shells for 2 to 3 hours. We then performed long-duration time-lapse imaging of the tissue samples, combining bright-field microscopy with a fluorescence readout of the Ca²⁺ activity, enabling simultaneous tracking of morphology and tissue-wide Ca²⁺ dynamics. Movies were acquired at 1 min intervals over many hours, providing a faithful record of recurrent inflation and deflation cycles during regeneration for multiple tissue samples. Under all the perturbations described below, fragments completed regeneration into mature animals within approximately 48 h. We focus here on the early stages of the regeneration process, before significant morphological changes occurred[28]. During this period, the projected area exhibited a characteristic sawtooth trace, with slow increases interrupted by sharp drops (Fig. 1) that we use below as a quantitative proxy for pressure-release events.

We extract the projected area $A(t)$ from bright-field images and quantify changes using a fixed 1 min increment. This choice is motivated by a clear separation of timescales: deflation typically occurs on timescales shorter than 1 min, whereas swelling is much slower. For each sample, we compute the discrete difference $\Delta A(t) = A(t + 1 \text{ min}) - A(t)$, and focus on negative increments, $\Delta A(t) < 0$. The event size is defined as the absolute area drop at the end of the 1 min interval, $\Delta A_k = |\Delta A(t_k)|$. For comparisons across samples, we normalize event sizes by the mean area size of each tissue, defining $x = |\Delta A|/\langle A \rangle_t$, where $\langle \cdot \rangle_t$ denotes an average over all frames in the time trace of that tissue (see Methods). Negative area differences can arise from sources other than lumen deflation, including tissue contraction and elongation, rotation, motion toward or away from the objective, and segmentation noise. However, the tail of the distribution is dominated by genuine, sizeable deflation events; on the 1 min timescale, contributions from other sources are typically small. This is particularly true in the early stages of regeneration analyzed here: the lumen fluid is effectively incompressible, so as long as overall morphological changes remain modest, they do not significantly affect the enclosed volume. Also, small reorientations or mild deformations produce only minor fluctuations in the projected area. By contrast, rupture-driven deflation changes the enclosed volume abruptly, generating the large negative area increments that populate the tail of the distribution.

To probe how repair efficiency shapes area-drop statistics, we modulate the Ca²⁺ activity by utilizing two perturbations that act in opposite directions, as established in our previous works[28-30]: *Heptanol*, a small alcohol molecule that suppresses Ca²⁺ activity by disrupting intercellular coupling via gap junctions, and the application of a moderate-amplitude electric field that excites the epithelial tissues and elevates Ca²⁺ activity. Regeneration proceeds into a normal mature *Hydra* under both conditions. They allow us to compare event statistics across regimes of reduced versus enhanced Ca²⁺ activity using the same imaging and analysis pipeline.

Figures 2A–C show the complementary cumulative distributions (1 – CDF) of the normalized event sizes $x$ under three experimental conditions. Under a normal condition, the tail is dominated by an exponential cutoff, indicating a characteristic drop scale (Fig. 2A). Applying an external electric field shifts this cutoff to a somewhat smaller value, consistent with stronger suppression of large deflation events (Fig. 2B). In contrast, reduced Ca²⁺ activity by the blocked gap junctions with *Heptanol* yields a markedly heavier tail, consistent with an approximate power-law regime over the accessible tail range

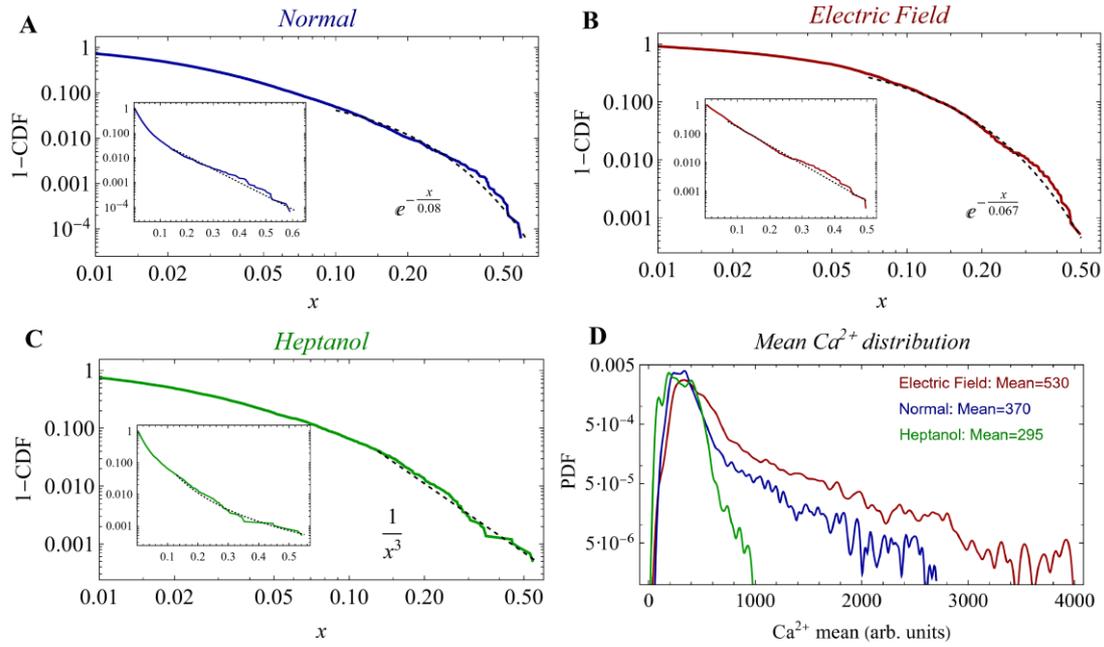

**Figure 2 | Ca²⁺-linked statistics of projected area drops. (A–C)** Complementary cumulative distribution functions, 1–CDF, of projected area-drop magnitudes under normal conditions (A), applied electric field (B), and under *Heptanol* (C). The normalized projected-area time traces for all independently regenerating tissues included in these statistics are shown in Supplementary Figs. S1–S3. Drop sizes are normalized as $x = |\Delta A|/\langle A \rangle_t$, where $\langle A \rangle_t$ denotes the mean projected area averaged over all frames in the time trace of a tissue; data are then pooled across tissues within each condition. Main panels are shown on log–log axes; insets show the same data on semi-log axes to emphasize the tail structure. Under reduced Ca²⁺ activity (*Heptanol*) the distribution exhibits a long tail consistent with a power-law behavior, whereas under normal conditions and with an applied electric field, the tail is dominated by an exponential cutoff, with the electric field slightly shifting the characteristic drop scale to smaller values. Dashed lines show tail fits: in (A) and (B), the tails are fit to $a\exp(-x/c)$, yielding $c = 0.083$ with 95% confidence interval [0.082,0.085] for (A) and $c = 0.068$ with 95% confidence interval [0.067,0.069] for (B). In (D), the dashed line shows a fit to $a/x^3$; allowing the exponent to vary in a fit to $a/x^b$ gives an optimal value $b = 3.1$ with 95% confidence interval [3.05, 3.14]. (D) Distributions (semi-log) of the spatial mean Ca²⁺ activity for the same type of samples as in (A–C), computed per frame by averaging over the tissue region and pooled over time[27]; the legend reports the mean Ca²⁺ level for each condition. Data are compiled from independent regenerating tissue fragments; the number of tissues is 24, 6, and 6 for (A), (B), and (C), respectively.

(Fig. 2C). Because we focus on the asymptotic tail, which is populated by the largest negative area changes associated with genuine pressure-release events, these differences reflect rare-event statistics independent of the analysis conventions. The emergence of a power-law tail under *Heptanol* signals a qualitative change toward more marginal, critical-like failure dynamics as we explain below.

To relate these statistics to the conjunctly measured Ca²⁺ activity we utilize a fluorescence probe in our strain (see Methods)[27, 30]. Fig. 2D shows the distribution of spatially averaged fluorescence signals reflecting the Ca²⁺ activity (computed per frame over the entire tissue region) for each condition. The ordering of the fluorescence distributions mirrors the behavior of tails of the area-drop statistics: lower Ca²⁺ activity coincides with heavier-tailed event statistics, whereas baseline or elevated Ca²⁺ activity is associated with an exponential cutoff. This correlation is consistent with impaired Ca²⁺ signaling, particularly the reduced intercellular propagation when gap junctions are blocked, weakening or delaying the contractile response that arrests rupture growth.

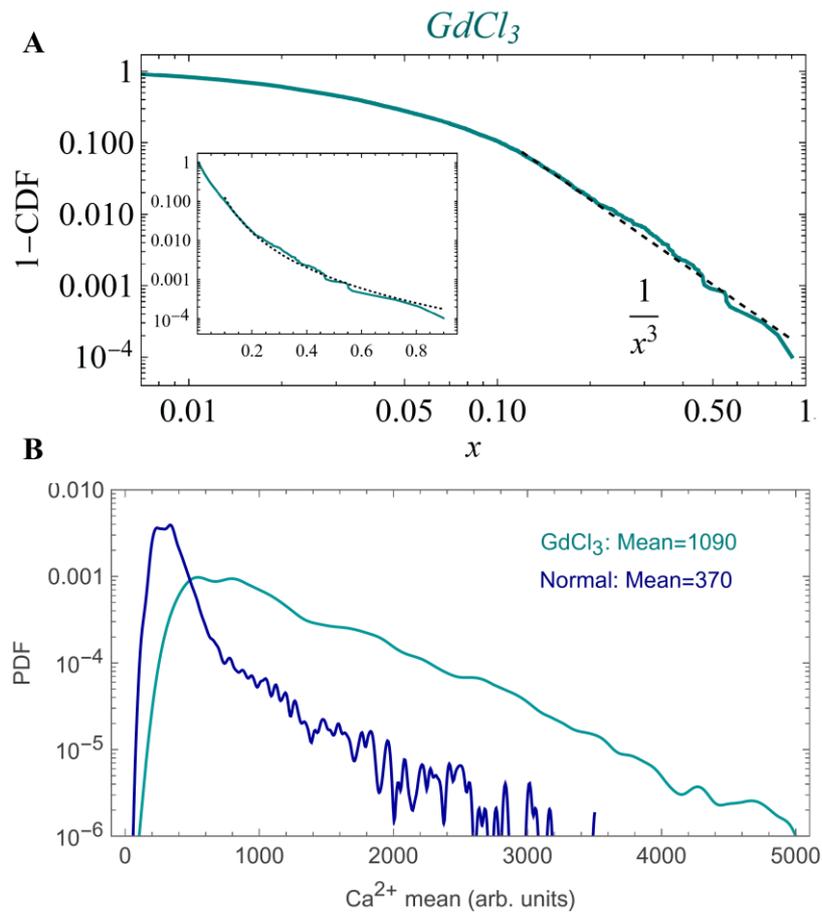

**Figure 3 | Disrupting the stretch-activated Ca²⁺ pathway enhances the tail of the event size distribution.** (A) Complementary cumulative distribution, 1−CDF, of normalized projected area-drop, $x = |\Delta A|/\langle A \rangle_t$, under *GdCl₃* for 13 independent regenerating tissue fragments. Individual normalized projected-area time traces for these tissues are shown in Supplementary Fig. S4. The main panel is shown on log–log axes; the inset shows the same data on semi-log axes to highlight the tail structure. The dashed line indicates a fit of the form $a/x^3$ over the tail region; allowing the exponent to vary in a fit to $a/x^b$ gives an optimal value $b = 3.11$ with 95% confidence interval [3.06,3.17]. (B) Distributions (semi-log) of the spatially averaged Ca²⁺ activity, computed for each frame by averaging over the tissue region and pooled over time. Curves compare tissues treated with *GdCl₃* (the same tissues as in panel A) with untreated controls. The mean Ca²⁺ level for each condition is indicated in the panel.

A direct prediction from the strong sensitivity of rupture statistics to the Ca²⁺ response is that other perturbations that compromise mechanoresponsive calcium signaling, should produce a similar shift in the tail behavior. One such independent route to reducing resealing efficiency is to target the mechanosensitive Ca²⁺ pathway that couples tissue stretch to repair activation. Towards this end, we utilize *GdCl₃*, commonly used to inhibit stretch-activated channels, in particular the Piezo-mediated Ca²⁺ entry[31, 32]. This perturbation is expected to weaken and slow down the response triggered by abrupt changes in membrane tension and tissue stretching near a rupture[33-35]. Indeed, similar to the effect of *Heptanol*, *GdCl₃* produces a pronounced heavy tail in the drop-size distribution (Fig. 3A), consistent with an approximate power-law regime with a similar power as under *Heptanol*. These statistics reflect a stronger contribution of rare, large events compared with the baseline exponential cutoff (Fig. 2A). Importantly, this shift in the behavior of area-drops is not explained by a global reduction in the Ca²⁺ activity: the distribution of spatially averaged $Ca^{2+}$ signal under *GdCl₃* shifts to higher values than in the normal case (Fig. 3B). Thus, the control of the rupture–repair cycle needs not be set by the global $Ca^{2+}$ level, but by the efficiency of the local mechanochemical coupling at the rupture boundary: the conversion of abrupt stretch changes into a fast repair response at the rupture boundary. A straightforward interpretation is that inhibition of stretch-triggered $Ca^{2+}$ entry, which activates local actomyosin contractile forces, slows the response to rupture progression and allows the rupture front to advance further through the heterogeneous tissue and to encounter more local weak tissue connections and barriers.

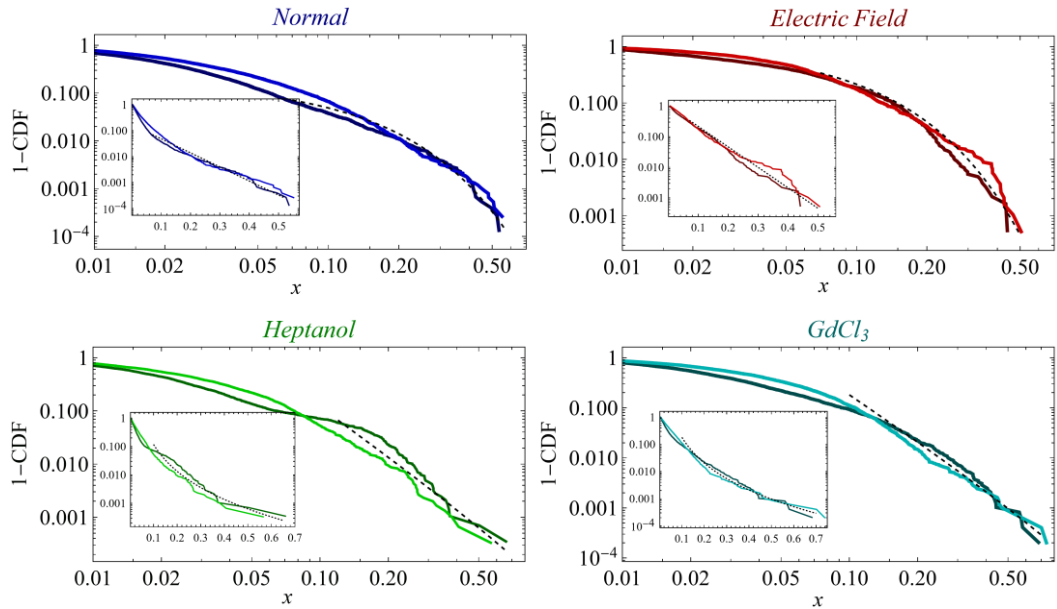

**Figure 4 | Stationarity of rupture event statistics.** Complementary cumulative distributions, 1−CDF, of normalized projected area-drop sizes computed separately for the early and late halves of each regeneration time series. The main panel is shown on log–log axes; the inset shows the same data on semi-log axes to highlight the tail structure. The time trace is analyzed, separately, for two equal-duration segments within the early stages of regeneration, before significant morphological changes. For each half, area-drop sizes are normalized by the mean projected area of that same half. Darker curves denote the early half and lighter curves the late half. The close overlap between early and late distributions indicates that the event-size statistics are approximately stationary over this stage. Panels show the same four conditions as in Figs. 2–3: normal, applied electric field, *Heptanol*, and *GdCl₃*.

An important question raised by the heavy-tail statistics in the area-drop distributions is whether they reflect a stationary rupture process or instead a drift, as the tissue progresses along the early stages of the regeneration trajectory before major morphological changes set in[19]. If repair efficiency or mechanical integrity systematically matures over time, the probability of the largest pressure-release events should change along the trace. To test this, we split each time series into an early and late half and compute the complementary cumulative distribution, 1−CDF, of area drop separately for each segment. Figure 4 shows the resulting distributions for all the cases considered above, with darker curves denoting the early half and lighter curves the late half. Across conditions, the two curves nearly overlap, indicating that event statistics remain essentially stationary throughout the regeneration trajectories considered here, with only minor differences between early and late periods.

**Model**

A simple model that explains the above results can be constructed according to three principles: rupture as an avalanche process in a heterogeneous load-bearing tissue, calcium-linked feedback that controls the effective proximity to marginal stability, and a mapping from the rupture extent to the measured projected area drop.

We treat the epithelial tissue as a mechanically heterogeneous sheet whose local resistance to opening varies in space and time[36]. Slow osmotic loading increases lumen pressure and thereby raises tensile stress in the sheet quasi-statically, whereas rupture and relaxation during a release event are fast. In this separation of timescales, a macroscopic rupture is expected to occur as a cascade: local yielding redistributes the stress along the tensioned sheet and can trigger additional yielding elsewhere[36, 37]. Such cascade (avalanche) dynamics are a generic consequence of slow driving in disordered load-bearing media and provide a natural basis for broad event-size statistics. The mechanisms that determine where rupture initiates in the tissue lie beyond the scope of this work and their discussion is not required for our central conclusion; for discussions of this issue, see Refs.[19, 20].

In the regime where the system repeatedly approaches marginal stability, mean-field fracture and branching arguments predict a broad avalanche-size distribution (see Supplementary Note for details) of the scaling form[38, 39]

$$p(s) \sim s^{-5/2} f(s/s_c), \qquad (1)$$

where $s$ is the avalanche size, $s_c$ is a cutoff scale, and $f(y)$ is a cutoff function that is approximately constant for $y \ll 1$ and decays rapidly for $y \gg 1$ [e.g. $f(y) = \exp(-y)$]. In this framework, $s_c$ contains the physics of active mechanical response to the tissue's tension and rupture. It is the key control parameter which sets the weight of rare, very large cascades and is the quantity most directly tuned by our perturbations.

The cutoff scale $s_c$ is set by the negative feedback that arrests a rupture cascade during stress release. Two feedback mechanisms act in parallel. First, pressure release due to the tissue's rupture provides passive stabilizing feedback: as the opening vents the lumen, the driving load drops tending to slow down further growth. Second, Ca²⁺-linked forces provide active feedback: mechanically evoked Ca²⁺ signals recruit actomyosin contractility, reinforce new adhesion junctions and promote wound closure. By arresting the advancing tissue's tear and shortening the time window over which the lumen can vent, this active response suppresses cascade growth beyond what passive unloading alone can achieve. In the language of branching processes, the combined feedback keeps the effective branching ratio well below unity, so the dynamics remain strongly subcritical, $s_c$ stays small, and event sizes retain a characteristic scale with an exponentially decaying tail. When the Ca²⁺-linked repair is compromised, either by disrupting intercellular signaling with *Heptanol* or by inhibiting stretch-activated Ca²⁺ entry with *GdCl₃*, the active feedback is weakened, and rupture propagation operates closer to marginality. As a result, an extended avalanche regime emerges, visible as a power-law tail in the event-size statistics; the tail is truncated only at very large events by the finite pressure release during venting, which imposes a high cutoff scale, $s_c$.

To connect the avalanche size $s$ to the experimentally measured area drop $\Delta A$, it is useful to introduce the associated pressure drop $\Delta P$, defined as the difference between the lumen pressure just before rupture and immediately after resealing, and consider how $\Delta P$ depends on $s$. Because the projected area is a smooth function of pressure, a linear expansion around the pre-rupture state implies that the event-induced changes are proportional, $\Delta A \propto \Delta P$.

The magnitude of $\Delta P$ is set by how much fluid is discharged before the tissue reseals[40]. In a hydraulic discharge picture[41], this discharge is controlled by two factors: the effective hydraulic conductance $G$ of the opening and the effective time $\tau$ for which it remains open, giving $\Delta P \propto G\tau$. We expect $G$ to scale approximately linearly with the effective open area created during the rupture event, $s$: for a given lumen-to-exterior pressure difference, the total venting flux is the sum of local fluxes across the open region, so increasing rupture area adds flow paths in parallel and yields $G \propto s$.

The scaling of the open time $\tau$ follows by treating resealing as a nucleation-limited process occurring along the boundary of the opened rupture region. Repair activity is initiated stochastically anywhere along the rupture boundary, with an approximately constant rate per unit boundary length. Larger events, therefore, provide larger targets for initiating the closure process. If the effective open area created during an event is $s$, the associated boundary length scales as $\ell \propto \sqrt{s}$, so the total nucleation rate for initiating closure somewhere along the boundary grows in proportion to $\ell$. The expected waiting time to the first successful nucleation event is then $\tau_{\text{wait}} \propto 1/\ell$. If the subsequent closure process, once nucleated, is fast compared with $\tau_{\text{wait}}$, the open time is dominated by nucleation, yielding $\tau \approx \tau_{\text{wait}} \propto 1/\sqrt{s}$.

Taken together, these scaling arguments yield a direct link between the measured drop size and rupture extent. Using $\Delta A \propto \Delta P$, $\Delta P \propto G\tau$, $G \propto s$, and $\tau \propto 1/\sqrt{s}$, we obtain $\Delta A \propto \sqrt{s}$. Substituting $s \propto \Delta A^2$ into Eq. (1) then gives a corresponding distribution for area drops, $p(\Delta A) \propto \Delta A^{-4} \exp(-|\Delta A|/A_c)$, where the cutoff $A_c$ is set by the underlying cutoff $s_c$ and thus by the repair efficiency. In practice, the finite number of events makes direct estimation of the probability density $p(\Delta A)$ sensitive to binning choices, especially in the tail. We therefore analyze the complementary cumulative distribution, $1-\text{CDF}$, of the normalized drop size $x = |\Delta A|/\langle A \rangle_t$, which is more robust in the rare-event regime. Under the mapping derived above, it yields

$$1 - \text{CDF} \sim \frac{1}{x^3} \exp\left(-\frac{x}{x_c}\right) \qquad (2)$$

where $x_c$ is the cutoff set by repair efficiency. When $x_c$ is large, the cutoff is pushed to high $x$, and the tail is effectively a power law: $1-\text{CDF} \sim x^{-3}$. This behavior is observed under *Heptanol* (Fig. 2C) and *GdCl₃* (Fig. 3A), conditions that compromise $Ca^{2+}$-linked resealing.

**Conclusions**

*Hydra* regeneration from a tissue fragment commonly exhibits repeated episodes of pressure build-up due to osmotic gradients and its release by a tissue rupture. This, in turn, requires rapid repair to support the proper continuation of morphogenesis and the maintenance of cellular homeostasis. Tissue rupture is therefore fundamentally different from the commonly observed material failures in physical systems. To quantify these events, we represent each pressure-release episode by its projected-area drop and analyze the resulting event-size distribution. The tail of this distribution provides the most sensitive probe of the rupture–repair dynamics: under baseline conditions, it shows a characteristic scale with an exponential cutoff, whereas weakening of the resealing mechanism exposes power-law heavy tails consistent with a critical-like regime of stress release.

A mean-field avalanche picture reproduces the power-law tail observed experimentally when the repair process is weakened. In this framework, rupture growth is treated as a cascade in a heterogeneous landscape, with a size distribution given by Eq. (1). The central control parameter is the cutoff $s_c$, which sets the probability of rare, very large cascades. By introducing a hydraulic discharge link between pressure drop and venting, together with a nucleation-limited resealing time that scales with rupture boundary length, we obtain a mapping from rupture extent to the measured area drop. This mapping predicts that, on one hand, when the effective cutoff is pushed to large values, the measured tail becomes, effectively, a power-law, as observed under *Heptanol* and *GdCl₃*. On the other hand, when the cutoff remains small, large events are suppressed, and the distribution exhibits an exponential tail with a characteristic event scale.

Our data indicate that the tissue remains subcritical throughout the regeneration process and runaway ruptures, although in principle possible[20], do not occur under our experimental conditions. Nevertheless, one expects that, depending on the *Hydra* strain, tissue heterogeneity, and swelling rate, the system may dynamically reach a pressure that exceeds the critical threshold, at which a macroscopic runaway rupture becomes likely (see Supplementary Note 1). In that case, the tissue is expected to fold back into a closed shell over much longer times, in a process resembling the folding of an excised flat tissue to a closed shell at the early stages of regeneration. Such a process is consistent with the prolonged healing times reported in Ref. [20]. No such runaway ruptures were observed in our experiments (Supplementary Figs. S1–S4), and all the tissue samples analyzed here regenerated into a mature *Hydra* within a normal time window.

Our perturbations identify Ca²⁺-linked repair as a dominant active control of this cutoff. Two independent interventions reduce resealing efficiency: disrupting intercellular coupling with *Heptanol* and inhibiting stretch-activated Ca²⁺ entry with *GdCl₃*. Both shift the statistics toward heavier tails. Together, these observations identify mechanically evoked Ca²⁺ signaling as the dominant feedback that arrests rupture growth. Pressure release provides passive unloading, but the reproducible tail shifts under these perturbations show that active repair largely sets the cutoff of large rupture episodes.

The same results also point to the importance of tissue heterogeneity. The emergence of heavy tails when repair is slowed down is consistent with rupture advancing intermittently through a spatially varying mechanical landscape, sampling local weak spots and barriers. In this view, repair does more than restore cohesion. It controls how deeply rupture dynamics can explore the tissue heterogeneity, and therefore whether stress release remains confined to a characteristic scale or develops broad, critical-like fluctuations. The approximate stationarity of the statistics before a major morphological transition occurs[28] further supports treating this behavior as a well-defined dynamical regime, rather than a slow drift along the regeneration trajectory.

More broadly, these findings place rupture and resealing in a regenerating *Hydra* tissue, alongside a wide class of front propagation and avalanche phenomena in driven disordered systems. In such systems, event statistics reflect the competition between slow loading, heterogeneous barriers, and feedback that limit cascade growth[42, 43]. Examples include earthquakes, brittle fracture, plastic deformation, and Barkhausen noise. These systems are characterized by power-law statistics emerging near marginal stability, where the cutoff encodes the strength of stabilization[42-44]. Our results provide a biological counterpart, where stabilization does not reflect pure material or geometrical properties, but is actively regulated through mechanosensitive signaling and contractile repair. This crucial difference between these physical examples and living tissues reflects the participation of active forces triggered by the $Ca^{2+}$ signaling, which supports repeated episodes of rupture-repair, and securing the maintenance of the tissue's integrity over time. In this view, repair efficiency acts as a control axis that shifts rupture statistics toward or away from critical-like behavior, allowing flexibility and adaptability under changing conditions; the hallmark of a living tissue.

**Materials and Methods**

*Experimental Methods*
All experiments are carried out with a transgenic strain of *Hydra Vulgaris* (*AEP*) carrying a GCaMP6s fluorescence probe for $Ca^{2+}$ (see Refs. [27, 28, 30] for details of the strain). Animals are cultivated in a *Hydra* culture medium (HM; 1mM NaHCO3, 1mM CaCl2, 0.1mM MgCl2, 0.1mM KCl, 1mM Tris-HCl pH 7.7) at 18°C. The animals are fed every other day with live *Artemia nauplii* and washed after ~4 hours. Experiments are initiated ~24 hours after feeding. Small tissue fragments are excised from different regions along the axis of a mature *Hydra*. A thin ring is first cut from each of these regions and is further cut into (usually) 4 small fragments. These tissue fragments are incubated in a dish for ~3 hrs to allow their folding into spheroids prior to transferring them to the experimental sample holder under flow of HM.

For the tissue samples regenerated under *Heptanol*, the flowing medium is replaced with HM containing 300 μl/l of *Heptanol* (1- Heptanol, 99%, Alfa Aesar) after good mixing[28]. Tissue samples regenerated under an external electric field are observed in the same HM medium under a constant 30V AC field at 2 kHz (4 mm distance between the electrodes)[28]. The experimental area traces for samples under normal conditions, *Heptanol* and electric field analyzed in this work, are the same as those presented in Refs[6, 27, 28]. For the experiments with *GdCl₃*: 1M stock (Sigma 439770 powder dissolved in water) was diluted to a final concentration of 50μM in HM medium and introduced to the experimental setup under the same flow conditions throughout the experiment as above. Some comparison experiments were performed with higher concentrations of *GdCl₃* (100-150 μM) showing similar results for the area-drop statistics and were therefore combined with the lower concentration data.

The experimental setup is similar to the one described before[6] (for details of the experimental electric field setup see Refs[28-30]). In all the experiments, spheroid tissues are placed within wells of ~1.3 mm diameter made in a strip of 2% agarose gel (Sigma) to keep the regenerating *Hydra* in place during time-lapse imaging. The tissue spheroids are free to move within the wells. In all the experiments a peristaltic pump (IPC, Ismatec, Futtererstr, Germany) flows the medium continuously from an external reservoir (replaced at least once every 24 hrs) at a rate of 170 ml/hr into two channels around the samples. All the experiments are done at room temperature.

*Microscopy*
Time-lapse bright-field and fluorescence images are taken by a Zeiss Axio-observer microscope (Zeiss, Oberkochen Germany) with a 5× air objective (NA=0.25) and a 1.6× optovar and acquired by a CCD camera (Zyla 5.5 sCMOS, Andor, Belfast, Northern Ireland). The sample holder is placed on a movable stage (Marzhauser, Germany), and the entire microscopy system is operated by Micromanager, recording images at 1 min intervals. The recording at 1 min resolution allows faithful analysis of the area drops over long experiments, without tissue damage, and recordings from multiple tissue samples

*Data Analysis*
For the analysis, images are reduced to 696x520 pixels (~1.6 μm per pixel) using ImageJ. Masks depicting the projected tissue shape are determined for a time-lapse movie using the bright-field (BF) images by a segmentation algorithm described in [45] and a custom code written in Matlab. Shape analysis of regenerating *Hydra*'s tissue is done by representing the projected shape of the tissue by polygonal outlines as presented before[46]. The fluorescence analysis is done on images reduced to the same size as the bright-field ones (696x520 pixels).

*Normalization of event size*

To motivate the dimensionless drop $x=|\Delta A|/\langle A\rangle_t$, we use a thin-shell scaling picture. For an inflated spherical shell of radius $R$ and thickness $h$, the in-plane stress, $T$, scales as $T\sim PR/h$, where $P$ is the lumen pressure relative to the exterior. A local failure tolerance $T_c$ sets a characteristic pressure scale $P_c \sim T_c\, h/R$. Because the tissue is heterogeneous and evolves during regeneration, a given rupture releases a random fraction of this scale, $\Delta P=\eta P_c$, with $\eta$ a dimensionless event-to-event factor. For small deformations, the projected area varies smoothly with pressure, and the fractional change satisfies $\Delta A/A\simeq\kappa\Delta P$, where the effective areal compliance scales as $\kappa\propto R/(K_{\text{eff}}h)$ with $K_{\text{eff}}$ as the effective tissue stiffness. Combining these scaling leads to $\Delta A/A\propto\eta\, T_c/K_{\text{eff}}$, i.e., the leading dependence on overall tissue size cancels, while event-to-event variability reflects the heterogeneous mechanical landscape and repair dynamics. This allows to pool events for the statistical analysis across samples. It also reduces sensitivity to small variations in imaging across experiments.

**Acknowledgements:**

We thank Shimon Marom, Yitzhak Rabin, Kinneret Keren, Jay Fienberg and Omri Gat, for useful discussions and comments on the manuscript. EB thanks Liora Garion for technical help. This work was supported by a joint grant (OA & EB) from the Israel Science Foundation (Grant No. 1586/25).

**List of Supplementary Materials:**

Supplementary Figs. S1 to S5
Supplementary Notes 1
References

Supplementary Material for

# Rupture-Repair Cycles in Regenerating *Hydra* Tissues

Oded Agam[1] and Erez Braun[2]

[1]The Racah Institute of Physics, Edmond J. Safra Campus, The Hebrew University of Jerusalem, Jerusalem 9190401, Israel.

[2] Department of Physics and Network Biology Research Laboratories, Technion-Israel Institute of Technology, Haifa 32000, Israel.

**This PDF file includes:**

Supplementary Figs. S1 to S5
Supplementary Notes 1
References

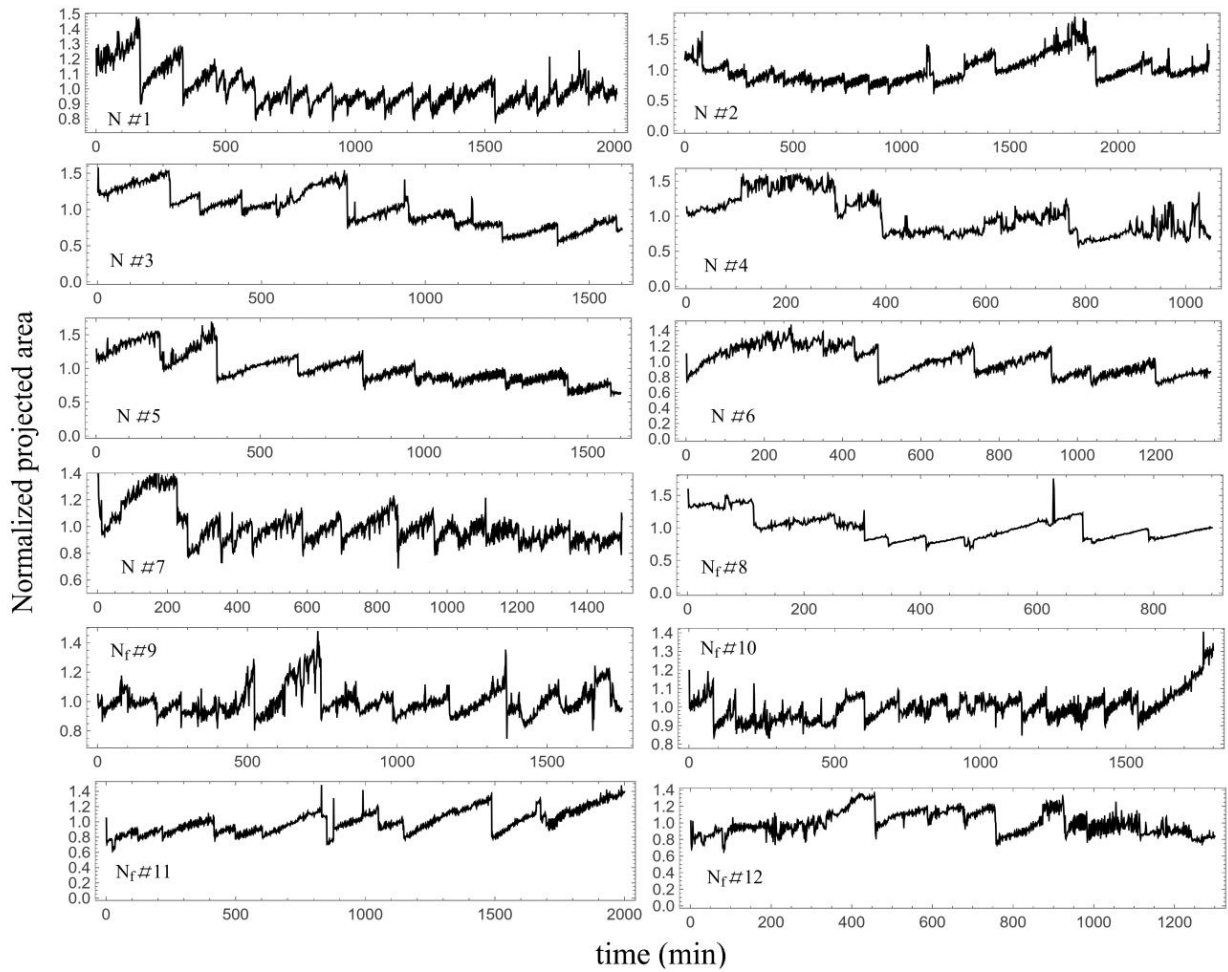

**Fig. S1 | Time traces of the normalized projected area for individual tissues under normal conditions.** Time is measured from the first imaging time point (2–3 h after excision) and traces are truncated before the onset of major morphological changes. For each tissue, the projected area is normalized by its mean over the displayed time trace.

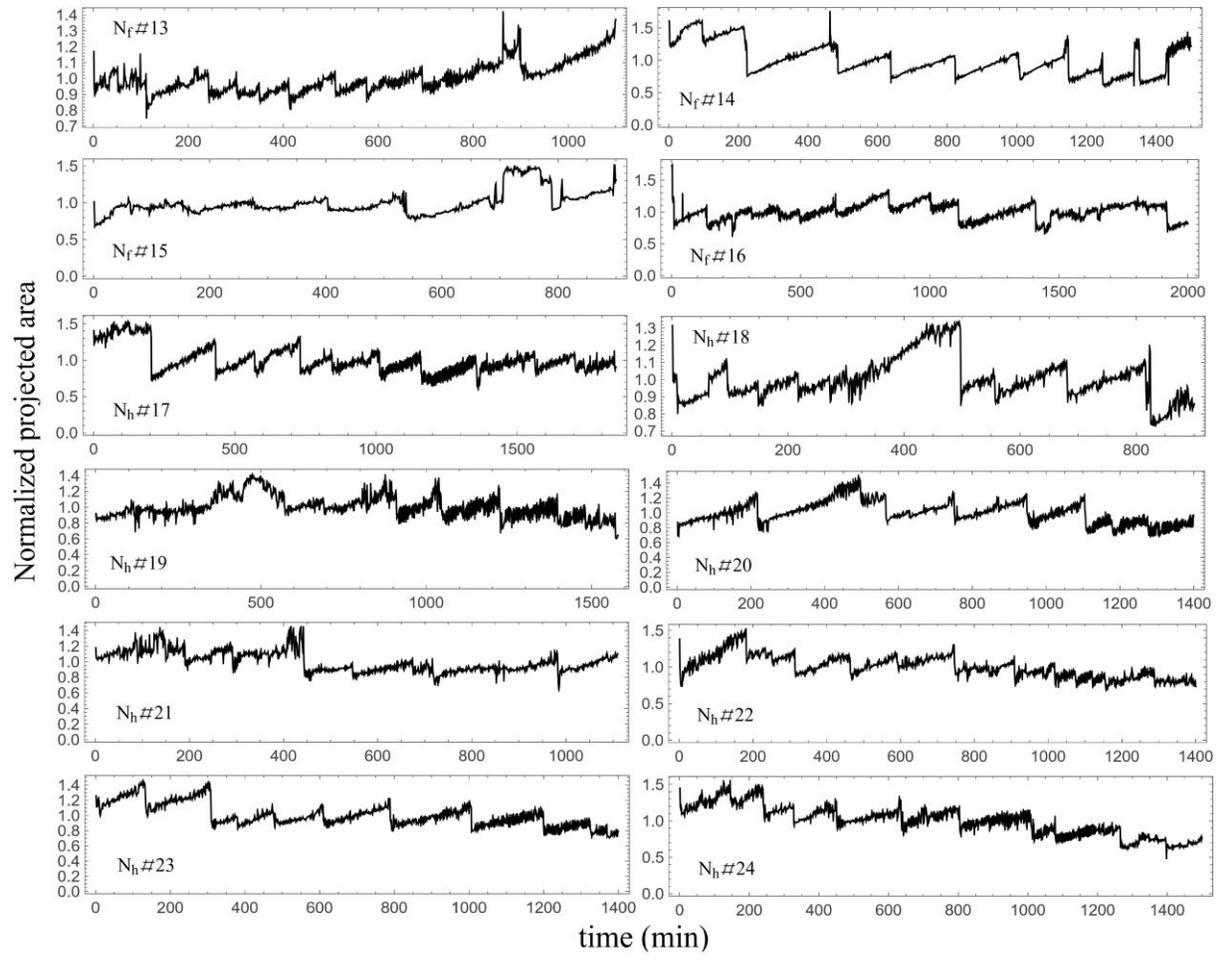

**Fig. S1 (Cont.)**

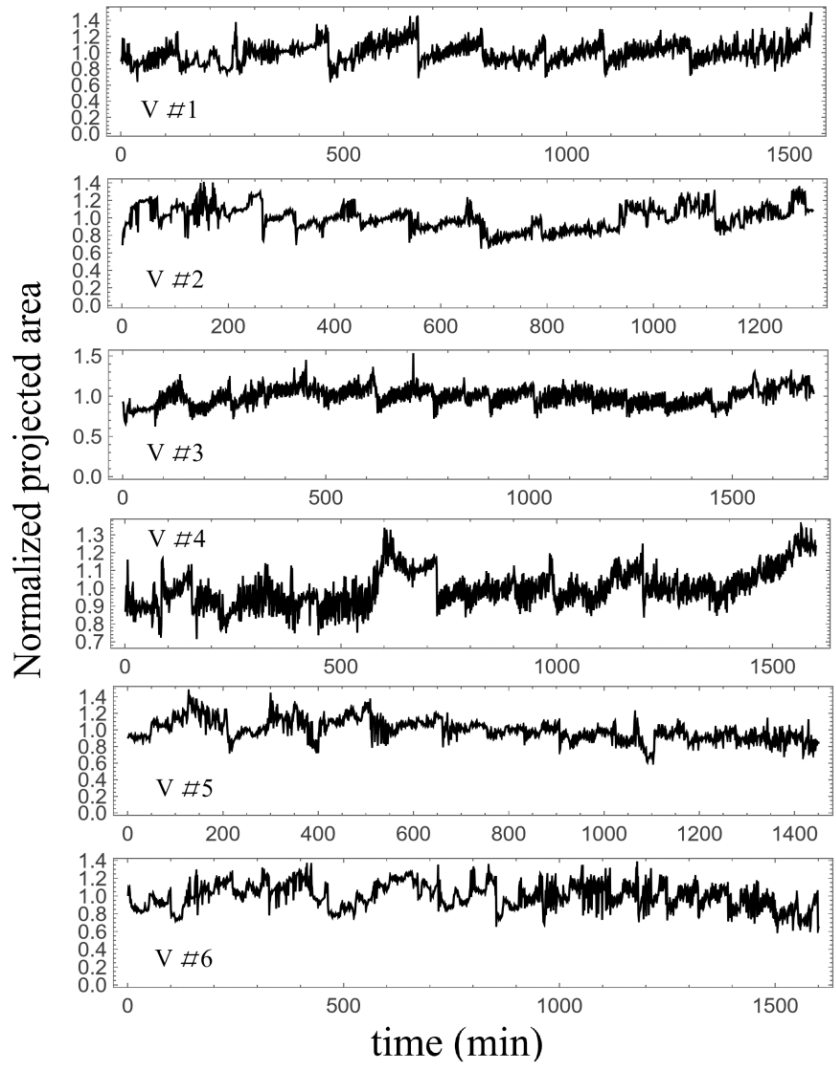

**Fig. S2| Time traces of the normalized projected area for individual tissues subjected to electric field.** Time is measured from the first imaging time point (2–3 h after excision) and traces are truncated before the onset of major morphological changes. For each tissue, the projected area is normalized by its mean over the displayed time trace.

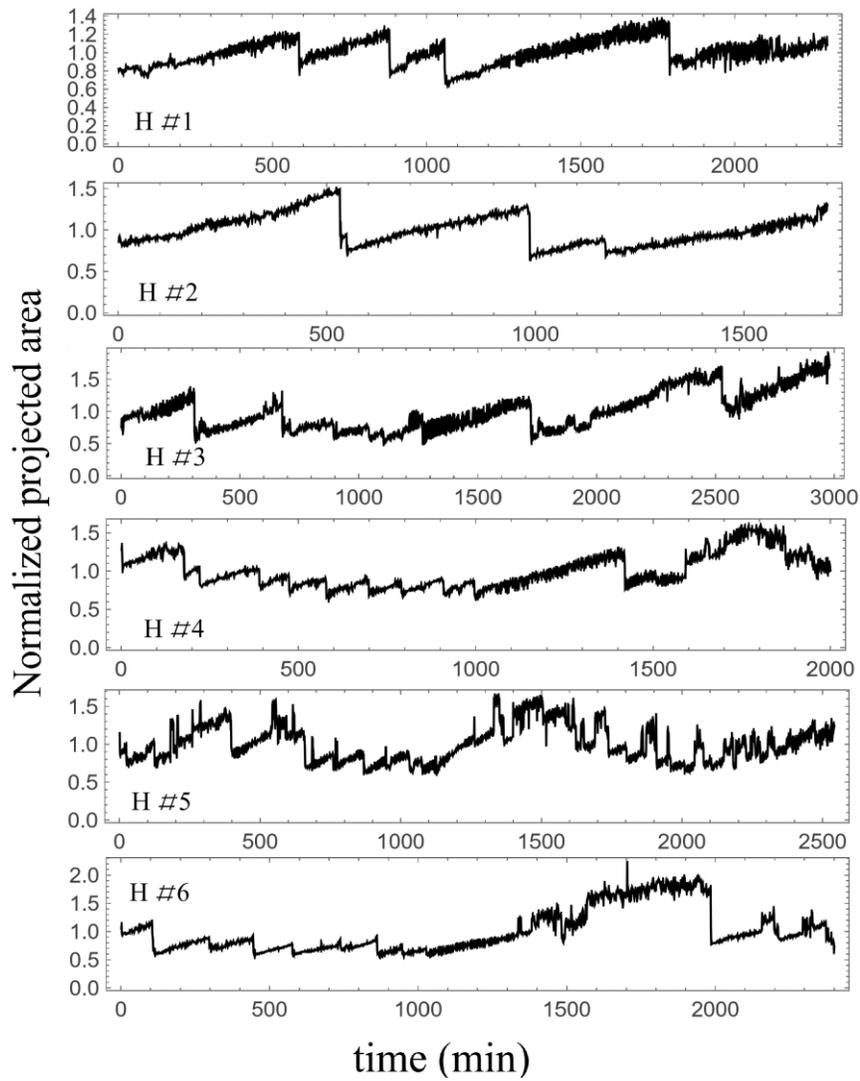

**Fig. S3| Time traces of the normalized projected area for individual tissues administrated with *Heptanol*.** Time is measured from the first imaging time point (2–3 h after excision) and traces are truncated before the onset of major morphological changes. For each tissue, the projected area is normalized by its mean over the displayed time trace.

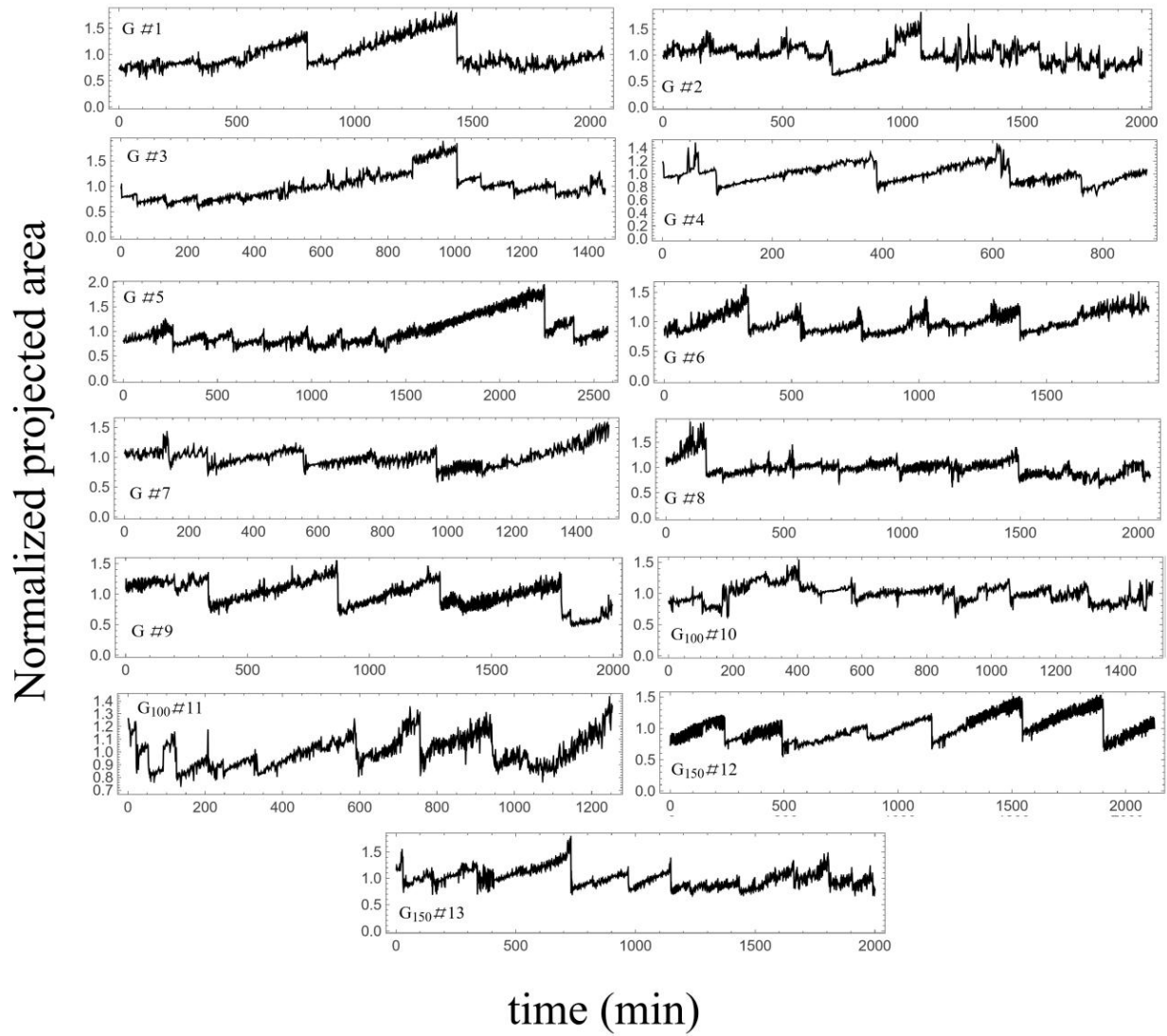

**Fig. S4| Time traces of the normalized projected area for individual tissues administrated with *GdCl₃*.** Time is measured from the first imaging time point (2–3 h after excision) and traces are truncated before the onset of major morphological changes. For each tissue, the projected area is normalized by its mean over the displayed time trace.

**Supplementary Note 1: Mean field derivation of rupture avalanche statistics**

In this note, we derive the mean field prediction for the distribution of rupture event sizes in a heterogeneous, slowly driven tissue. Our goal is to obtain the conditional law[1, 2]

$$p(s) \sim s^{-5/2} \exp(-c\varepsilon_*^2 s), \tag{S1}$$

where $s$ is the rupture size (interpreted as the effective opened area in a coarse-grained description), $\varepsilon_*$ measures the effective distance from marginal instability, and $c$ is a system-dependent constant. The derivation proceeds in three stages. First, we formulate rupture propagation as a mean-field avalanche in a disordered sheet under quasi-static loading by lumen pressure. Second, at fixed proximity to instability, we map the avalanche to a biased random walk and obtain the standard conditional result[3, 4]

$$p(s|\varepsilon) \sim s^{-3/2} \exp(-c\varepsilon^2 s). \tag{S2}$$

Finally, because the system is slowly driven, rupture events do not all start at the same value of ϵ, which marks the distance from marginality. The gradual pressure buildup brings the tissue progressively closer to marginal propagation, so the ensemble of events is weighted toward small ϵ, and this additional weighting shifts the mean field exponent from -3/2 to -5/2[5].

In the first stage, the key idea is that rupture growth can be viewed as an avalanche process[6]. The rupture boundary (tear front) propagates through a tissue that is mechanically heterogeneous, meaning that different locations along the front resist opening by different amounts because of variations in junctional integrity, cytoskeletal organization, and local geometry[7]. We coarse-grain the front into mesoscopic segments on the scale of this heterogeneity and define an elementary event as a local advance of the front by one segment, which increases the opened area by a typical amount. Such a local advance does not remain local in its mechanical consequences: because the epithelial sheet is under tension[8, 9], advancing the front redistributes stress along the boundary and in the surrounding sheet, changing the driving force on other segments[6, 7, 10]. Some of those segments may have been close to their local failure thresholds, and the stress redistribution can push them over the threshold, triggering additional front advances[11]. These secondary advances can, in turn, trigger further advances, producing a cascade of correlated micro advances that collectively constitute a single macroscopic rupture episode[6, 11].

In a mean-field description, we do not track where triggering occurs along the boundary; instead, we replace the detailed, geometry-dependent stress redistribution with its average statistical effect and characterize cascades by an offspring number, defined as the number of additional elementary boundary advances triggered by a given advance 12. This approximation is justified here for two related reasons. First, the lumen pressure provides a spatially global drive: it, together with the actomyosin forces[10], sets the in-plane tension of the closed sheet, thereby biasing all boundary segments toward opening. Second, in a tensioned epithelial sheet, a local front advance redistributes stress along the sheet and can influence many other boundary sites, not only immediate neighbors, so an elementary event typically couples weakly but broadly to the rest of the boundary. In this regime, the statistics of total event size are controlled primarily by the average triggering strength rather than by the detailed spatial pattern of who triggers whom[14].

The mean offspring number, averaged over disorder realizations and elementary advances at fixed loading, $m = \langle K \rangle$, quantifies how close the system is to runaway propagation: $m < 1$ corresponds to subcritical propagation where cascades remain finite, whereas $m=1$ is the marginal point at which the front becomes critical[12, 13]. This motivates writing $m=1-\varepsilon$, with $\varepsilon > 0$ measuring the distance from marginality In our system, $m$ is controlled by the lumen pressure, $P$: as pressure increases, tensile load rises, and a larger fraction of boundary segments sit close to threshold, so the probability that one advance triggers others increases and $m(P)$ approaches one from below. During the slow inflation stage before rupture, $P(t)$ increases quasi-statically and $\varepsilon[P(t)]$ decreases toward zero. A rupture episode initiates when a local seed event occurs at some pressure $P_0$ corresponding to an initial $\varepsilon = \varepsilon(P_0)$ that is typically small.

However, the pressure is not constant during the rupture episode. As soon as an opening forms, fluid vents and the lumen pressure decreases, relaxing the stress. Since $m(P)$ is an increasing function of pressure, venting causes $m$ to decrease during the event. This is an intrinsic negative feedback that helps terminate propagation. The branching description remains consistent provided there is a time scale separation: the time $\delta t$ for an elementary boundary segment to fail is short compared with the time $t_P$ over which pressure changes appreciably due to venting, so $\delta t \ll t_P$. Then the rupture consists of many fast elementary advances occurring while $P(t)$ and hence $\varepsilon(t)$, drifts. Imaging does not resolve $\delta t$, so the macroscopic area drop can appear instantaneous on the minute time scale while still containing many unresolved micro advances.

However, our experiments indicate that this hydraulic stress relaxation is not the dominant factor setting the observed cutoff. The key evidence is that the cutoff scale can be tuned effectively by manipulating the efficiency of repair, through calcium-linked activity, while the osmotic loading and venting physics remain present under all conditions. This implies that, in the regime we probe, rupture growth is typically arrested primarily by active resealing at the boundary rather than by pressure decay alone. In other words, venting contributes to a background stabilization, but the experimentally controllable process that limits the branching of the rupture front and sets the effective cutoff is the active repair mechanism affected by the local $Ca^{2+}$ level and response.

For completeness, we now briefly sketch the standard derivation of the conditional probability given by Eq. (S2)[12, 13]. The derivation uses the classic mapping between the total number of events in a near-critical branching cascade and a first-passage problem for a biased random walk[15]; this result is well known in the branching-process and avalanche literature (see, e.g., standard probability texts[16] and reviews on mean-field avalanche statistics[13]).

Assume that at fixed $\varepsilon$, the offspring numbers $\{K_n\}$ are independent and identically distributed with mean $\langle K \rangle = 1-\varepsilon$ and finite variance $\sigma^2 = \langle K^2 \rangle - \langle K \rangle^2$. To compute the event-size distribution, imagine a list of elementary advances whose triggered consequences have not yet been generated. Initially, the list contains only the seed advance, and at step $n$, you remove one element from the list, draw its number of triggered advances $K_n$, and append those $K_n$ new advances to the list. $Z_n$ is the length of that list after $n$ removals. Thus $Z_0 = 1$ and the cascade ends when there are no pending steps left, i.e. when $Z_n = 0$. Thus, when we explore one active event we remove it, contributing $-1$, and add its offspring $K_n$. This gives the recursion

$$Z_{n+1} = Z_n - 1 + K_n. \tag{S3}$$

Let us define the increments $X_n = K_n - 1$. Then the drift is $\langle X \rangle = -\varepsilon$ while the variance is $Var(X) = \sigma^2$. Thus $Z_n$ is a biased random walk on the nonnegative integers, started at 1 and absorbed at 0, and $s$ is the first passage time to 0.

For large avalanches, we use a diffusion approximation and replace the discrete random walk by the continuous stochastic process[17]:

$$dZ = -\varepsilon dt + \sigma dW_t, \tag{S4}$$

where $dW_t$ is the infinitesimal increment of a standard Wiener process (Brownian motion). The process is defined with an absorbing boundary at $Z = 0$, and the initial condition $Z(0) = 1$. The first passage time density to hit zero for a drift diffusion process has the inverse Gaussian form[18]:

$$f(t) = \frac{1}{\sqrt{2\pi\sigma^2 t^3}} \exp\left(\frac{(1-\varepsilon t)^2}{2\sigma^2 t}\right), \tag{S5}$$

Identifying $t$ with the avalanche size $s$ up to microscopic time units yields $p(s|\varepsilon) \approx f(s)$ in the large $s$ limit, and expanding the exponent to the leading order in $1/s$ leads to Eq. (S2) with $c = 1/(2\sigma^2)$. This conditional law expresses two universal features: a $s^{-3/2}$ power law inherited from one dimensional first passage, and an exponential cutoff at $s^* \sim \varepsilon^{-2}$ introduced by the subcritical drift $\varepsilon > 0$.

We now incorporate the slow driving by pressure buildup, which means that the ensemble of rupture events does not occur at a fixed $\varepsilon$ [19]. Instead, each event is initiated at some pressure level, hence some $\varepsilon$, and the observed distribution is an average over $\varepsilon$ with a proper weight function[5]. In our setting, however, it is more natural to perform this average over the initiation pressure $P$, because this pressure is the control parameter that increases quasi-statically during the inflation phase. Thus, we write the event-size distribution as an average of the conditional law over the pressures at which events are triggered,

$$p(s) = \int dP w_P(P) p(s|P) \tag{S6}$$

where $w_P(P)$ is the event-weighted density of initiation during slow loading, while $p(s|P)$ is given by (S2).

To rewrite the above average in terms of $\varepsilon$, we relate the distance from marginal propagation to the distance from the critical pressure[20]. A convenient mean-field way to do this is to introduce the load-sharing picture along the rupture boundary[2]. At a given pressure $P$, the sheet tension imposes a total tensile load on the boundary. This total load is carried only by the subset of boundary segments that remain intact at that moment. Let $\gamma$ denotes the typical load borne by an intact segment and let $q(\gamma)$ be the fraction of segments that can sustain that load. Under global load sharing, the sustainable total load is proportional to the product $P(\gamma) \propto q(\gamma)\gamma$.

As $\gamma$ increases, $q(\gamma)$ decreases because more segments exceed their local thresholds. For any reasonable threshold distribution, $q(\gamma) \to 0$ for large $\gamma$, therefore $\gamma q(\gamma)$ typically rises at small $\gamma$ (because most segments are intact) and eventually falls at large $\gamma$ (because almost all segments have failed). In other words $P(\gamma)$ has a maximum at some $\gamma_c$, which defines the mean-field stability limit $P_c = P(\gamma_c)$. Expanding near this maximum gives a quadratic departure, $P_c - P \propto (\gamma - \gamma_c)^2$.

Next, we connect $\gamma - \gamma_c$ to $\varepsilon$. In the branching description, $\varepsilon = 1 - m$ measures how far the cascade is from marginal propagation. In mean field, this distance is controlled by how close the boundary is to the stability limit, which is set by $\gamma$; thus near criticality it is natural to take $\varepsilon \propto \gamma_c - \gamma$. Combining the two relations yields $P_c - P \propto \varepsilon^2$. This mapping implies that $dP \propto \varepsilon d\varepsilon$ Therefore, if the event-weighted density $w_P(P)$ is smooth and non-vanishing at $P_c$ the pressure average is given by:

$$p(s) \sim \int_{\varepsilon_*}^{\infty} d\varepsilon\, \varepsilon\, p(s \mid \varepsilon), \tag{S7}$$

where $\varepsilon$ is the effective cutoff determined by the tissue's repair efficiency, and for simplicity we extended the upper limit of the integral to infinity because the contribution from the tail is exponentially small. Substituting the conditional mean-field result (S2) then gives (S1).

The mapping between the control parameter $P$ and the distance-to-marginality variable $\varepsilon$ is analogous to the binodal–spinodal structure of an equilibrium binary-mixture phase diagram[21]. To clarify this analogy, consider the Gibbs free-energy density $g(\chi;\lambda)$ as a function of a scalar state variable $\chi$ (the analogue of composition) and a control parameter $\lambda$. As sketched in Fig. S5A, for fixed $\lambda$, the candidate states are the stationary points $\partial_\chi g(\chi;\lambda) \equiv g'(\chi;\lambda) = 0$, and their local stability is determined by the curvature: $g''(\chi;\lambda) > 0$ corresponds to a locally stable (metastable) minimum, whereas $g''(\chi;\lambda) < 0$ corresponds to a locally unstable stationary point.

The binodal is the coexistence boundary that separates a single-phase region from a two-phase region. When thermodynamic parameters (for example, temperature at fixed overall composition) are tuned across this boundary, the globally stable state switches to the phase with lower Gibbs free energy. However, the phase that becomes globally unfavored need not disappear at the binodal: it can persist as a local minimum of $g(\chi;\lambda)$ separated from the stable phase by a finite barrier. In this metastable regime, small perturbations are restoring and relax back to the minimum; decay therefore requires a finite-amplitude fluctuation that crosses the barrier, i.e., a bifurcation-like nonlinear instability.

If the parameters are moved further into the coexistence region, the barrier decreases and eventually vanishes at the spinodal, which marks the limit of metastability. Beyond the spinodal the local minimum no longer exists, and the state is linearly unstable, so arbitrarily small fluctuations grow. At the spinodal, the metastable minimum approaches the nearby saddle until the two stationary points coalesce and disappear in a fold. This loss of local stability is signaled by vanishing curvature at the endpoint, $g''(\chi_s;\lambda) = 0$, the linear restoring force is therefore lost, and infinitesimal perturbations grow beyond the spinodal[22].

Near the spinodal endpoint $(\chi_s, \lambda_c)$ the scaling between the control parameter and the distance to local stability follows directly from the fold geometry. Define the deviations $\delta\chi = \chi - \chi_s$, and $\delta\lambda = \lambda_c - \lambda$. At the spinodal one has $g'(\chi_s;\lambda_c) = 0$ and $g''(\chi_s;\lambda_c) = 0$, so the leading nontrivial expansion of the Gibbs free energy takes the universal fold form:

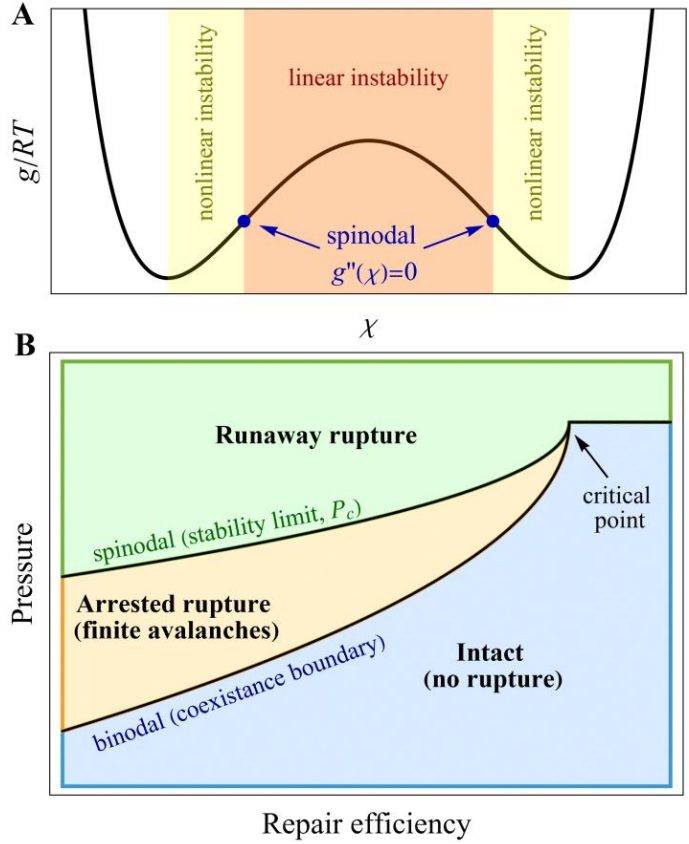

**Figure S5 | Binodal–spinodal analogy and schematic rupture "phase diagram".** (A) Gibbs free-energy landscape $g(\chi;\lambda)$ for a binary-mixture analogue, shown as a function of a scalar state variable $\chi$, and control parameter $\lambda$. Stationary points satisfy $\partial_\chi g(\chi;\lambda)=0$, and local stability is set by the curvature. The yellow regions indicate a metastable regime with $\partial_\chi^2 g(\chi;\lambda)>0$, where small perturbations are restoring and loss of the state requires a finite-amplitude fluctuation (nonlinear instability). The red region indicates linear instability with $\partial_\chi^2 g(\chi;\lambda)<0$, where infinitesimal perturbations grow. The spinodal points mark the boundary between these regimes and are defined by vanishing curvature, $\partial_\chi^2 g(\chi;\lambda)\big|_{\chi=\chi_s}=0$. (B) Corresponding mean-field rupture diagram in the plane of lumen pressure $P$, and repair efficiency (represented through the effective cutoff $\varepsilon_*$). Below a binodal-like boundary the tissue remains intact (no sustained rupture). Between the binodal-like boundary and the spinodal-like stability limit $P_c$, rupture can be initiated but remains arrested (finite avalanches followed by resealing). Above $P_c$, any threshold crossing triggers runaway rupture. A schematic critical point indicates where the arrested-rupture wedge closes at sufficiently high repair efficiency, in analogy with a critical endpoint of the equilibrium analogy.

$$g(\chi;P) \simeq g_s + \frac{u}{3}\delta\chi^3 - a\delta\lambda\delta\chi \tag{S8}$$

with $u>0$ and $a\neq 0$. Stationarity then gives $0=\partial_\chi g = u\delta\chi^2 - a\delta\lambda$, so the displacement of the stationary point obeys

$$\delta\chi^2 \propto \lambda_c - \lambda. \tag{S9}$$

The local stability is controlled by the curvature, $g''(\chi;\lambda) \simeq 2u\delta\chi$, and therefore it vanishes as $g'' \sim \sqrt{\delta\lambda}$ when $\lambda \to \lambda_c$.

Although the *Hydra*'s tissue considered here is active, driven and dissipative rather than thermodynamic, the mean-field rupture picture inherits the same fold-type stability structure as the equilibrium analogy. Here $P$ (lumen pressure) plays the role of the control parameter, while $\varepsilon$ quantifies the distance to marginal propagation, i.e., the linear stability of the arrested state. The spinodal expansion above implies that, near the endpoint where the arrested state disappears, the distance of the control-parameter from the stability limit is quadratic in the linear-stability measure. Namely, in analogy with Eq. (S8) we have $\varepsilon^2 \propto P_c - P$, and hence $dP \propto \varepsilon d\varepsilon$. Under quasi-static inflation, the system is slowly driven toward this spinodal-like stability limit, so rupture events naturally sample initiation pressures in its vicinity, leading to the shift of the mean-field exponent from $-3/2$ to $-5/2$.

Figure S5B summarizes the mean-field regimes in the plane of lumen pressure $P$ and repair efficiency. The lower boundary, labelled binodal, separates an intact region with no rupture from a region where rupture initiation becomes possible. Between this binodal-like boundary and the upper stability limit, labelled spinodal, rupture events are possible but remain arrested: threshold crossing produces finite avalanches (that eventually reseal). The curve $P_c$ as a function of repair efficiency marks the limit of existence of the arrested state in the mean field. For $P > P_c$, the system lies in the runaway-rupture region, where any threshold-crossing event leads to sustained propagation rather than resealing. At sufficiently high repair efficiency, the arrested wedge closes, and we indicate schematically a critical point at which the binodal-like and spinodal-like boundaries meet, in direct analogy with the termination of a coexistence region at a critical endpoint.